\documentclass[11pt]{article}
\usepackage{epsfig}

\newcommand{\EE}{\mbox{${\cal E}$}}

\newcommand{\FF}{\mbox{${\cal F}$}}

\newcommand{\eps}{\varepsilon}

\newcommand{\sfrac}[2]{{\textstyle\frac{#1}{#2}}}

\newcommand{\len}{\mathrm{len}}
\newcommand{\size}{\mathrm{size}}
\newcommand{\SMF}{\mbox{SMF}_n}
\newcommand{\bt}{\mathbf{t}}
\newcommand{\be}{\mathbf{e}}
\newcommand{\ff}{\mathbf{f}}
\newcommand{\bT}{\mathbf{T}}
\newcommand{\bpi}{\mbox{\boldmath$\pi$}}
\renewcommand{\root}{\mathrm{root}}
\begin{document}

\title{Percolation-like Scaling Exponents for Minimal Paths and Trees in
the Stochastic Mean Field Model}
\author{David J. Aldous\\Department of Statistics\\ 
367 Evans Hall \#\  3860\\
U.C. Berkeley CA 94720\\  aldous@stat.berkeley.edu}

\maketitle

\begin{abstract}
In the mean field (or random link) model there are $n$ points and
inter-point distances are independent random variables.
For $0 < \ell < \infty$
and in the $n \to \infty$ limit, 
let $\delta(\ell) = 1/n \times $
(maximum number of steps in a path whose average step-length is $\leq \ell$).
The function
$\delta(\ell)$ is analogous to the 
{\em percolation function}
in percolation theory: there is
a critical value $\ell_* = e^{-1}$ at which 
$\delta(\cdot)$ becomes non-zero, and
(presumably) a scaling exponent $\beta$ in the sense
$\delta(\ell) \asymp (\ell - \ell_*)^\beta$.
Recently developed probabilistic methodology
(in some sense a rephrasing of the cavity method 
developed in the 1980s by
M{\'e}zard and Parisi) 
provides a simple albeit non-rigorous way of writing down such functions in terms of solutions of 
fixed-point equations for probability distributions.
Solving numerically gives convincing evidence that $\beta = 3$.
A parallel study with {\em trees} and {\em connected edge-sets} in place of paths gives 
scaling exponent $2$,
while the analog for classical percolation has scaling exponent $1$. 
The new exponents coincide with those recently found in a different context 
(comparing optimal and near-optimal solutions of the mean-field 
TSP and MST problems), 
and reinforce the suggestion that scaling exponents determine
universality classes for optimization problems on random points.
\end{abstract}

\vspace{0.1in}


\newpage
\section{Introduction}
\label{sec-INT}
\subsection{Paths}
Consider $n$ points with inter-point distances
$(d(v,w) = d(w,v), 1 \leq v,w \leq n)$.
A {\em path} $\pi = (v_0,v_1,\ldots,v_m)$ 
visits a set of points, distinct except that maybe $v_m = v_0$.
Associated with a path $\pi$ is its length (number of steps)
$\len(\pi)$ and the average step-distance $A(\pi)$:
\begin{eqnarray*}
\len (\pi) &=& m \\
A(\pi) &=& m^{-1} \sum_{i=1}^m d(v_{i-1},v_i) .
\end{eqnarray*}
The celebrated
{\em Traveling Salesman Problem}
(TSP) concerns minimizing $A(\pi)$
subject to $\len(\pi) = n$.
One can also consider, for given $m <n$,
the question of the minimum value of $A(\pi)$ subject to $\len(\pi) \geq m$.
This has also been studied
as an algorithmic question 
\cite{arora97,balas89};
but instead we take a ``statistical physics" viewpoint of studying the
values $\min_\pi A(\pi)$ under a probability model for random points.
The most natural probability model is $n$ independent uniform random points
in the unit square, and study of the TSP in this model goes back
45 years to 
Beardwood et al
\cite{BHH59}.
See Steele \cite{steele97} for a recent survey of the general area.
Unfortunately the kind of questions we study seem far out of reach
of analytic methods in this two-dimensional model.
Instead we use a more tractable model with several names
(we say {\em stochastic mean-field} ($\SMF$) but also called
{\em random link} or {\em complete graph with random (exponentially distributed) edge-lengths})
which we imagine roughly as random points in
{\em infinite}-dimensional space.
Section \ref{sec-PWIT} provides details of the $\SMF$ model.
In the mid 1980s
M{\'e}zard and Parisi \cite{MP86} studied the TSP 
(and other optimization problems \cite{MP87,MPV87})
in the $\SMF$ model,
using the non-rigorous {\em cavity method} from statistical physics:
see \cite{MParisi03} for a recent survey of the cavity method.
Recent work of the author \cite{me94,me103,me101}
develops a methodology based on (additive) renormalization within an
infinite-point random model of distance.
This methodology, in some sense just a rephrasing of the cavity method, provides a consistent framework for a wide variety
of different calculations for different optimization problems in the
context of $\SMF$.

In this paper we study a deterministic function
$(\eps(\delta), 0 < \delta \leq 1)$
arising as the limit
\begin{equation}
\eps(\delta) = \lim_n 
E \min\{A(\pi): \ \len(\pi) \geq \delta n, \ \pi \mbox{ a path in } \SMF \} .
\label{def-eps}
\end{equation}
(Limits asserted here and later are presumed, but not rigorously
proved, to exist -- see section \ref{sec-method}.)
The value $\eps(1) \approx 2.04$
(obtained by numerically solving a fixed-point equation)
goes back to
M{\'e}zard and Parisi \cite{MP86},
while the value $\eps(0+) = e^{-1} \approx 0.368$
is given in Aldous \cite{me75} Proposition 7
(other aspects of paths are treated by
Janson \cite{janson-123}).
Our purpose is to show how 
the recent methodology
enables one to determine numerically the whole function $\eps(\delta)$.
A plot of the whole function is given in Figure 1 (left side).
Of particular interest is the scaling behavior as $\delta \downarrow 0$.
The numerical evidence (right side of Figure 1, and Table 1)
strongly suggests a scaling exponent 
\begin{equation}
\mbox{
$\eps(\delta) - \eps(0+) \asymp \delta^\alpha$
with $\alpha = 1/3$.
}
\label{path-scaling}
\end{equation}
This kind of scaling exponent is precisely analogous to scaling exponents
around the critical value in percolation theory,
as explained in section \ref{sec-analogy}.

\subsection{Trees}
There are parallel questions using trees in place of paths.
Consider a complete graph on $n$ vertices whose edges $e$
have lengths $d(e)$.
For any tree $\bt$ in the graph, with edges $e_1,\ldots,e_m$, write
$\size (\bt)$ for the number of edges of $\bt$ and 
$A(\bt)$ for the average edge-length:
\begin{eqnarray*}
\size (\bt) &=& m\\
A(\bt) &=& m^{-1} \sum_{e \in \bt} d(e) .
\end{eqnarray*}
The {\em Minimum Spanning Tree} (MST)
problem asks for the minimum of $A(\bt)$ subject to $\size (\bt) = n-1$.
Take $n$ random points in our stochastic mean field model
$\SMF$.
Analogously to the results for paths, we anticipate a deterministic function
$(\eps^*(\delta), 0 < \delta \leq 1)$
arising as the limit
\begin{equation}
\eps^*(\delta) = \lim_n 
E \min\{A(\bt): \ \size(\bt) \geq \delta n, \ \bt \mbox{ a tree in } \SMF \} .
\label{def-eps*}
\end{equation}
A well known result of Frieze \cite{fri85} for the MST says that
$\eps^*(1) = \zeta(3) \approx 1.202$,
whereas Aldous \cite{me75} argued 
$\eps^*(0+) \approx 0.263$
by numerics with fixed point equations.
Parallel to the study of paths,
our methodology
tells how
in principle
to determine numerically the whole function $\eps^*(\delta)$.
In practice we have not be able to carry this through
(see section \ref{sec-tree})
but instead have analyzed the following related question.
Instead of trees we consider
{\em connected edge-sets} 
$\be = (e_1,\ldots,e_m)$.
\newpage
\psfig{figure=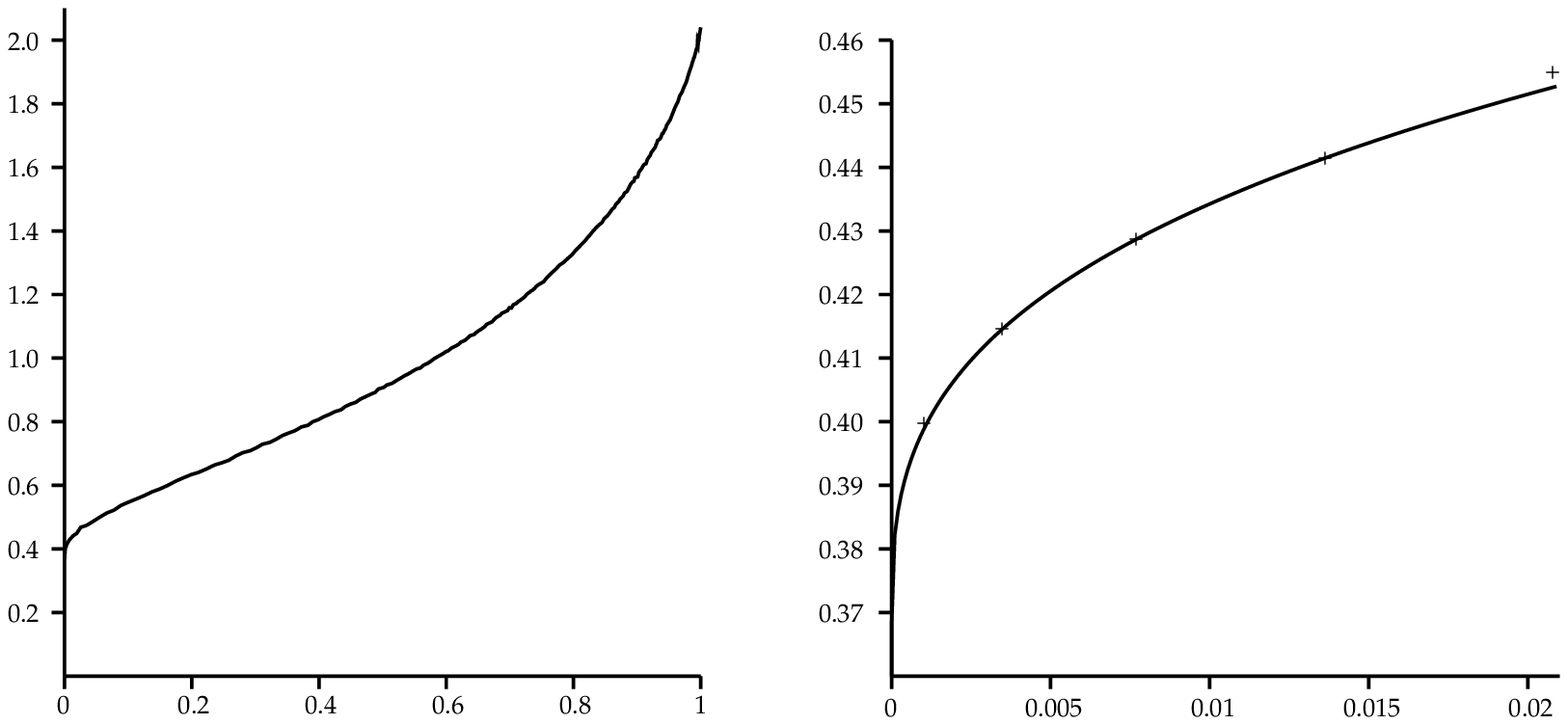}

\vspace{-0.2in}

{\bf Figure 1. 
The limit function for paths.}
{\small 
On the left is the function 
$\eps(\delta)$ defined at (\ref{def-eps}).
The horizontal axis is $\delta$,
the vertical axis is $\eps$.
The right side gives a close-up of the behavior for small $\delta$:
the points $+$ are the values estimated numerically in Table 1,
and the curve is
$\eps - e^{-1} = 0.308\  \delta^{1/3}$.
}

\vspace{0.7in}

\psfig{figure=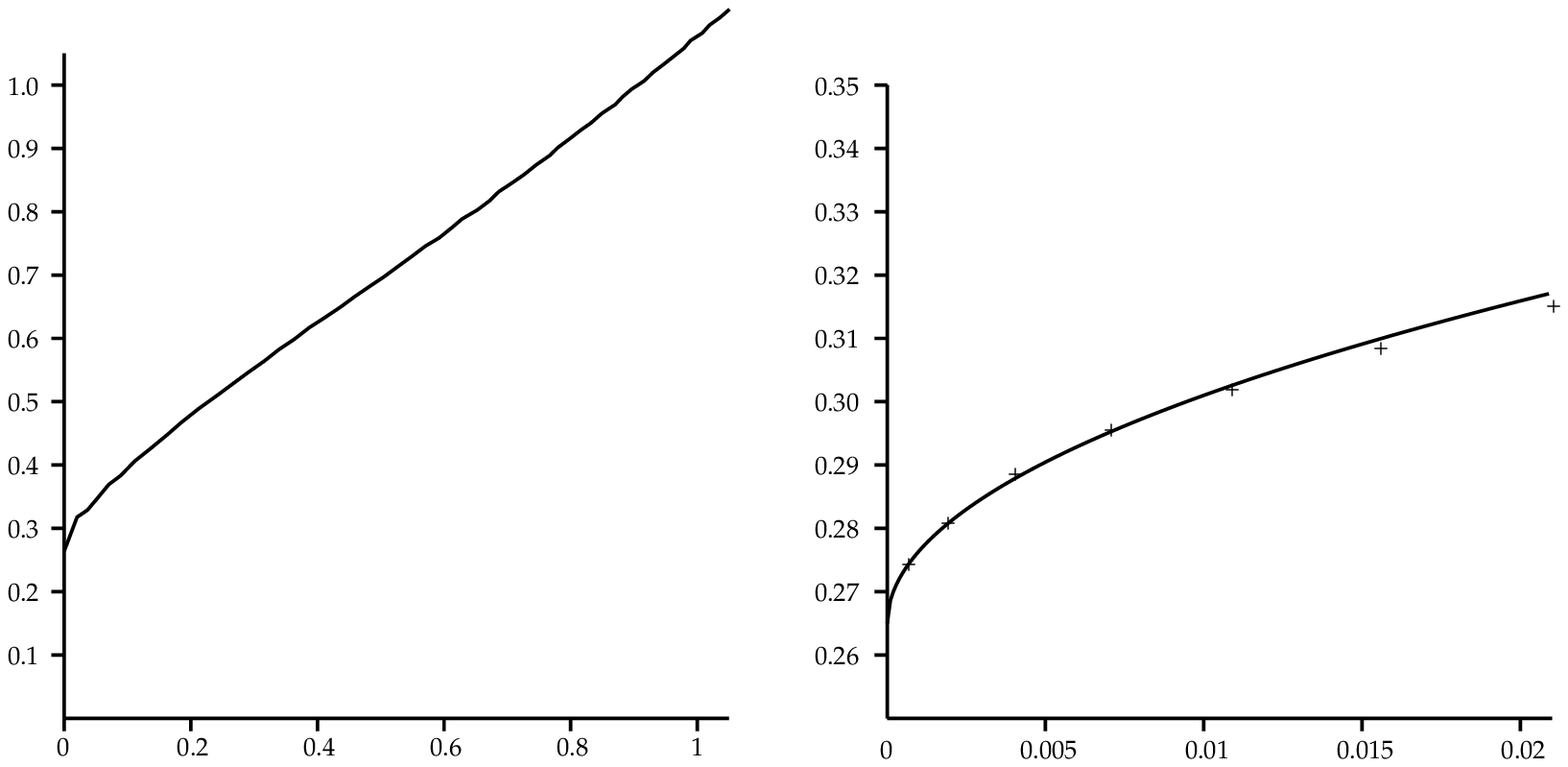}

\vspace{-0.2in}

{\bf Figure 2. 
The limit function for connected edge-sets.}
{\small 
On the left is the function 
$\tilde{\eps}(\delta)$ defined at (\ref{def-epstil}).
The right side gives a close-up of the behavior for small $\delta$:
the points $+$ are the values estimated numerically in Table 1,
and the curve is
$\tilde{\eps} - 0.265 = 0.360\  \delta^{1/2}$.
}

\newpage
Define
$\size (\be)$ and $A(\be)$ as above, and as at
(\ref{def-eps*})  
we anticipate a deterministic function
$(\tilde{\eps}(\delta), 0 < \delta < \infty)$
arising as the limit
\begin{equation}
\tilde{\eps}(\delta) = \lim_n 
E \min\{A(\be): \ \size(\be) \geq \delta n, \ \be \mbox{ a connected edge-set in } \SMF \} .
\label{def-epstil}
\end{equation}
A plot of the whole function is given in Figure 2 (left side).
Again, scaling as $\delta \downarrow 0$ is of interest.
The numerical evidence (right side of Figure 2, and Table 1)
gives an estimate $\tilde{\eps}(0+) \approx 0.265$ and
strongly suggests a scaling exponent 
\begin{equation}
\mbox{
$\tilde{\eps}(\delta) - \tilde{\eps}(0+) \asymp \delta^{\alpha^*}$
with $\alpha^* = 1/2$.
}
\label{tree-scaling}
\end{equation}
As explained in section \ref{sec-tree},
we must have the same $\delta \downarrow 0$ behavior for the ``tree" function
$\eps^*(\cdot)$ as for the ``connected edge-set" function
$\tilde{\eps}(\cdot)$.

\vspace{0.3in}

\begin{tabular}{lllc||lllc}
\multicolumn{2}{c}{$\eps(\delta)$}&
\multicolumn{2}{c||}{paths}&
\multicolumn{2}{c}{$\tilde{\eps}(\delta) = \eps^*(\delta)$}&
\multicolumn{2}{c}{trees}\\
\hline 
\ $\lambda$&\ \ \ $\delta$&$\eps - e^{-1}$&$\frac{\eps - e^{-1}}{\delta^{1/3}}$
&$\lambda$&\ \ \ $\delta$&$\eps - 0.265$&$\frac{\eps - 0.265}{\delta^{1/2}}$\\
0.53&.0386&.112&.332&0.34&.0211&.0502&.346 $\pm .001$\\
0.51&.0293&.100&.324&0.33&.0156&.0436&.349 $\pm .001$\\
0.49&.0209&.0872&.317&0.32&.0110&.0371&.355 $\pm .002$\\
0.47&.0137&.0737&.308&0.31&.00714&.0307&.364 $\pm .003$\\
0.45&.00773&.0610&.308&0.30&.00411&.0237&.370 $\pm .005$\\
0.43&.00352&.0468&.308&0.29&.00198&.0160&.360 $\pm .010$\\
0.41&.00108&.0321&.313&0.28&.000730&.0094&.348 $\pm .027$
\end{tabular}

\vspace{0.1in}
\noindent
{\bf Table 1.}
{\small 
Scaling behavior near the critical point, for
$\eps(\delta)$ (left side)
and
$\tilde{\eps}(\delta)$ (right side).
In each case the function is defined implicitly
via functions $\eps(\lambda)$ and $\delta(\lambda)$, 
as explained 
below (\ref{eps-lam}).
See section \ref{sec-sample} for discussion of the $\pm$ sampling error.
}

\subsection{The analogy with percolation functions}
\label{sec-analogy}
Instead of the functions $\eps(\delta)$ and $\tilde{\eps}(\delta)$ at
(\ref{def-eps},\ref{def-epstil}),
we could equivalently study their inverse functions
$\delta(\ell)$ and $\tilde{\delta}(\ell)$ whose interpretations are
\begin{eqnarray*}
\delta(\ell) &=& \lim_n 
E \max\{n^{-1} \  \len(\pi):\  A(\pi) \leq \ell, \ \pi \mbox{ a path in } \SMF \} .
\\
\tilde{\delta}(\ell) &=& \lim_n 
E \max\{n^{-1} \  \# \be:\  A(\be) \leq \ell,
 \ \be \mbox{ a connected edge-set in } \SMF \} .
\end{eqnarray*}
Of course the scaling exponent for trees at (\ref{tree-scaling})
can be rewritten as
\[
\tilde{\delta}(\ell) \asymp (\ell - \ell_*)^\beta
\mbox{ with } \beta_{\mbox{tree}} = 2
\]
for $\ell > \ell_* = \tilde{\eps}(0+)$.
Similarly the scaling exponent for paths at (\ref{path-scaling})
can be rewritten as
\[
\delta(\ell) \asymp (\ell - \ell_*)^\beta
\mbox{ with } \beta_{\mbox{path}} = 3
\]
for $\ell > \ell_* = \eps(0+) = e^{-1}$.
To make the analogy with percolation, 
for $0<t<\infty$ 
consider the maximal size connected edge-subset 
$\mathrm{perc}_n(t)$
such that 
\[ \max_{e \in \mathrm{perc}_n(t)} d(e) \leq t . \]
So $\mathrm{perc}_n(t)$
is the largest percolation cluster, that is the largest 
connected component of the subgraph of $\SMF$ consisting of edges of
length $\leq t$.
Well known theory concerning giant components in the random graph
process implies
\[ \lim_n n^{-1} E \# \mathrm{perc}_n(t) = p(t) \]
where $p(t)$ has the properties
\[ p(t) = 0, \ 0 \leq t \leq 1; \quad 
p(t) \sim 2(t-1) \mbox{ as } t \downarrow 1 . \]
Thus the scaling exponent for ordinary percolation in $\SMF$ is
$\beta_{\mbox{perc}} = 1$.
Note we can rewrite $p(\cdot)$ as
\[ p(\ell) = \lim_n 
E \max\{n^{-1} \  \# \be:\  \max_{e \in \be} d(e) \leq \ell,
 \ \be \mbox{ a connected edge-set in } \SMF \} .
\]
This differs from $\tilde{\delta}(\ell)$ only in the use of
$\max_{e \in \be} d(e)$
in place of
$\mathrm{ave}_{e \in \be} d(e)$.
So we have a rather precise analogy between our function and the
usual percolation function.

\subsection{The big picture}
This paper provides some pieces of a big picture.
Time is not yet ripe for a complete survey, but let us provide
some glimpses of other pieces.
Our main results here are the scaling exponents
$\beta_{\mbox{tree}} = 2, \ \beta_{\mbox{path}} = 3$
near the ``percolative critical values" 
$\eps^*(0+), \ \eps(0+)$.
In Aldous and Percus \cite{me103} we study a different notion
of ``scaling exponent" dealing with behavior near the ``spanning constants", 
i.e. near the
MST and TSP constants 
$\eps^*(1), \ \eps(1)$.
These exponents are based on comparing near-optimal solutions to the optimal
solution, and turn out to take the values 
$2$ and $3$.
These values hold in the $\SMF$ model by
the methodology used here,
and there is evidence (from Monte Carlo simulations) they hold for random points
in real $d \geq 2$ dimensional space.
That the ``percolative" scaling exponents in this paper coincide with 
the
``spanning" exponents of \cite{me103} is remarkable, and reinforces the idea
put forward in \cite{me103} that these scaling exponents provide
a natural way of defining ``universality classes" of optimization problems
on random points.
A natural next project is to study via Monte Carlo these percolative scaling exponents
for random points in $d \geq 2$ dimensions, although this seems algorithmically
difficult.
At the time of writing, the only one of the four exponents we understand
non-computationally is the tree/spanning exponent ``2", which is easily
explained \cite{me103} using the greedy algorithm for finding the MST.
See section \ref{sec-scope} for further remarks.

\subsection{Methodology}
\label{sec-method}
Here is our methodology, in brief.
\begin{itemize}
\item The stochastic mean field model for $n$ points
has a $n \to \infty$ limit, the PWIT (section \ref{sec-PWIT}).
\item Introducing Lagrange multipliers turns the
constrained maximization problem into an unconstrained
maximization problem.  One can formulate
the corresponding maximization problem for the PWIT,
and define random variables 
$(X,Y)$ 
measuring the relative effect on the maximized value of including 
or excluding a reference edge in the solution.
\item The recursive structure of the PWIT enables one
to write down equations
(\ref{X-rec},\ref{Z-rec})
satisfied by $(X,Y)$, which can be numerically solved.
The limit optimal values of length and $A(\cdot)$ are
determined from the definitions of $(X,Y)$.
\end{itemize}
The arguments are not mathematically rigorous, for two main reasons.
First, the central idea of identifying limits of
solutions of
finite-$n$ optimization problems with 
solutions of 
infinite-$n$ optimization problems 
requires justification, which has been given only in the
case (related to but slightly different from those considered
here) of mean-field
{\em minimal matching} \cite{me94}
and the less closely related case of some random graph problems
\cite{GNS04}.
Second, the scaling exponents are found by numerically solving
equations with a parameter and examining numerical behavior
as the parameter goes to a limit, and this falls short of 
analyzing the parameter-limit behavior rigorously.

\section{The stochastic mean field model and its infinite-point limit}
\label{sec-PWIT}
For fixed $n$, the $\SMF$ model is defined as follows.
There are $n$ points.
For each of the ${n \choose 2}$ pairs of points, there is
a ``link" whose length is random with exponential (mean $n$)
distributions, these random lengths being independent.
The distance between two points is then the length of the shortest
path of links between them.
The assumption of {\em exponential} distribution is convenient
but not essential; results are unchanged if the link lengths
are $nL$ where $L>0$ has a density with $f_{L}(0+) = 1$.

The scaling of link lengths is set up so that, as $n \to \infty$,
the mean distance from a typical point to its nearest neighbor
converges to $1$.
But much more is true, as we now outline briefly
(see \cite{me101} for detailed survey).
There is an infinite-point model, the {\em PWIT},
defined as follows.  
There is a root $\emptyset$.
The root has an infinite number of links to points labeled
$(1,2,3,\ldots)$, and these link lengths
$0<\xi^\emptyset_1 < \xi^\emptyset_2 < \ldots$
are the successive points of a Poisson process of rate $1$
on $(0,\infty)$.
Recursively, each point $i$ has an infinite number of further
links to points $(i1,i2,i3,\ldots)$ whose lengths
$0<\xi^i_1 < \xi^i_2 < \ldots$
are independent copies of the Poisson process.
The PWIT is illustrated in Figure 3, and the web site
\cite{mePWITweb} enables one to explore its structure via
genuine simulations.

The PWIT is the $n \to \infty$ limit of $\SMF$ in a precise
sense called 
{\em local weak convergence} \cite{me101}.
Choose a random point of $\SMF$ to be a root.
Then as $n \to \infty$, for any fixed ``window size" $r$
the configuration of points in $\SMF$ within a window of radius
$r$ centered at the root
converges in distribution to
the configuration of points in the PWIT within a window of radius
$r$ centered at the root.

Two properties of the PWIT enter into our calculations later.

{\bf (a).} 
For each ``child" $i$ linked to the root, there is a subtree
$\bT_i$ consisting of $i$ and its descendants.
The {\em recursive structure of the PWIT},
built into the definition, says that the subtrees $\bT_i$
are independent as $i$ varies and are distributed as the PWIT itself.

{\bf (b).} 
The fact that we choose a (uniform) {\em random} vertex of
$\SMF$ to be the root leads to a {\em stationarity} property
of the PWIT.
Roughly, this says that the root is a ``typical"
vertex of the PWIT and therefore, by the ergodic principle,
we can compute averages over all vertices of the PWIT
by computing expectations at the root.
As a more explicit instance,
given a random vertex subset $A_n$ of $\SMF$, suppose we
have joint local weak convergence of
$(\SMF,A_n)$ to $(\mathrm{PWIT},A_\infty)$
for a random vertex subset $A_\infty$ of the PWIT.
Then $n^{-1} E \# A_n \to P(\root \in A_\infty)$, 
where \# denotes cardinality.
Note that here $A_n$ is dependent on 
$\SMF$, but the root of $\SMF$ is then chosen independently of $A_n$.

\setlength{\unitlength}{0.3in}
\begin{picture}(12,12)
\thinlines
\put(0.5,0.5){\line(1,0){11}}
\put(0.5,0.5){\line(0,1){11}}
\put(11.5,11.5){\line(0,-1){11}}
\put(11.5,11.5){\line(-1,0){11}}
\put(6,6){\circle*{0.2}}
\put(6,5){\circle*{0.2}}
\put(6,5.8){\line(0,-1){0.6}}
\put(6.1,5.3){$\xi^{\emptyset}_1$}
\put(4,8){\circle*{0.2}}
\put(5.8,6.2){\line(-1,1){1.6}}
\put(4.7,6.6){$\xi^{\emptyset}_3$}
\put(8,7){\circle*{0.2}}
\put(6.2,6.1){\line(2,1){1.6}}
\put(7.0,6.3){$\xi^{\emptyset}_2$}
\put(11,5){\circle*{0.2}}
\put(6.2,5.96){\line(5,-1){4.6}}
\put(8.5,5.1){$\xi^{\emptyset}_4$}
\put(5,4){\circle*{0.2}}
\put(5.8,4.8){\line(-1,-1){0.6}}
\put(7,2){\circle*{0.2}}
\put(6.07,4.8){\line(1,-3){0.86}}
\put(10,1){\circle*{0.2}}
\put(6.2,4.8){\line(1,-1){3.6}}
\put(4,2){\circle*{0.2}}
\put(4.9,3.8){\line(-1,-2){0.8}}
\put(5,1){\circle*{0.2}}
\put(5,3.8){\line(0,-1){2.6}}
\put(3,8){\circle*{0.2}}
\put(3.8,8){\line(-1,0){0.6}}
\put(1,10){\circle*{0.2}}
\put(2.8,8.2){\line(-1,1){1.6}}
\put(4,10){\circle*{0.2}}
\put(4,8.2){\line(0,1){1.6}}
\put(7,11){\circle*{0.2}}
\put(4.2,8.2){\line(1,1){2.6}}
\put(2,4){\circle*{0.2}}
\put(1,5){\circle*{0.2}}
\put(1.8,4.2){\line(-1,1){0.6}}
\put(3.9,7.8){\line(-1,-2){1.8}}
\put(2,2){\circle*{0.2}}
\put(2,3.8){\line(0,-1){1.6}}
\put(9,10){\circle*{0.2}}
\put(8.07,7.2){\line(1,3){0.86}}
\put(10,8){\circle*{0.2}}
\put(8.2,7.1){\line(2,1){1.6}}
\put(11,8){\circle*{0.2}}
\put(10.2,8){\line(1,0){0.6}}
\put(11,11){\circle*{0.2}}
\put(10.07,8.2){\line(1,3){0.86}}
\put(5.6,5.9){$\emptyset$}
\put(4.17,7.94){$\scriptstyle{3}$}
\put(8.16,6.84){$\scriptstyle{2}$}
\put(5.61,4.89){$\scriptstyle{1}$}
\put(10.61,4.9){$\scriptstyle{4}$}
\end{picture}

{\bf Figure 3.  The PWIT.}
{\small 
Illustration of the vertices of the PWIT within a window
of radius $3$ centered on the root $\emptyset$.
Lines indicate the links, but are drawn only
when both end-vertices are within the window.
Thus the four links at $\emptyset$ shown are at distances
$0< \xi^{\emptyset}_1 < \xi^{\emptyset}_2 < \xi^{\emptyset}_3 < \xi^{\emptyset}_4 < 3$
from $\emptyset$, while there are an infinite number of
links at $\emptyset$ of lengths greater than $3$.
Orientation of lines in pictures is arbitrary.
}

\section{The recursive distributional equation: the path case}
\label{sec-RDE}
By introducing a Lagrange multiplier $\lambda > 0$,
the finite-$n$ problem of minimizing $A(\pi)$
subject to $\len(\pi)$ can be reformulated as
\begin{eqnarray*}
\mbox{\tt maximize }&:& \lambda \sfrac{\len(\pi)}{n} - A(\pi) \\
 \mbox{\tt subject to}&:&\mbox{ \ $\pi$ a path in $\SMF$}. \end{eqnarray*}
This has a random solution $\pi_n(\lambda)$.
We expect that as $n \to \infty$
\begin{eqnarray}
n^{-1} E  \len(\pi_n(\lambda)) &\to& \delta(\lambda)\label{del-lam}\\
EA(\pi_n(\lambda)) &\to& \eps(\lambda) \label{eps-lam}
\end{eqnarray}
and that the function $\eps(\delta)$
at (\ref{def-eps}) is determined implicitly via
the two functions
$\delta(\lambda), \eps(\lambda)$.

To set up the analogous optimization problem on the PWIT,
we first define what will be seen to be sets of feasible solutions.
Write $\bpi = (\pi_1,\pi_2,\ldots)$
for a family of vertex-disjoint doubly-infinite paths
in the PWIT.  
Define
\begin{itemize}
\item 
\ \ $\EE_0$ is the set of such families for which \underline{no path}
goes through the root;
\item 
\ \ $\EE_2$ is the set of such families for which \underline{some path}
goes through the root;
\item 
\ \ $\EE_1$ is the set of such families, 
where in addition to the doubly-infinite paths there 
exists exactly one singly-infinite path, and this path starts at the root.
\end{itemize}
Note the subscript indicates degree of root 
in the family.
For $\bpi = (\pi_u) \in \EE_0 \cup \EE_1 \cup \EE_2$
consider the objective function
\[
b(\bpi) = \lambda \times \#\{v: v \mbox{ a vertex of some } \pi_u\}
- \sum_{e: e \mbox{ edge of some } \pi_u} \xi_e
\]
with the convention that, for $\bpi \in \EE_1$, the root vertex
counts as $1/2$.
(Recall that $\xi_e$ is the length of edge $e$ in the PWIT.)
In the limit procedure which takes $\SMF$
to the PWIT, the limits of ``paths of length order $n$"
is exactly the set $\EE_0 \cup \EE_2$ of families of
doubly-infinite paths.
Thus the optimization problem on the PWIT can be written
symbolically as
\begin{equation}
\mbox{ maximize } b(\bpi) \mbox{ over } 
\bpi \in \EE_0 \cup \EE_2 .
\label{inf-prob}
\end{equation}
We seek to study the $\bpi$
that attains the maximum.
But we can't work directly with definition (\ref{inf-prob}),
because $b(\bpi)$ is the difference of two sums, each sum
having value $+ \infty$.
Instead we can consider 
{\em differences} between maximized $b(\cdot)$ values.
Specifically, given a realization of the PWIT we define
realizations of two random variables via
\begin{eqnarray}
X &=& \max_{\bpi \in \EE_1} b(\bpi)
- \max_{\bpi \in \EE_0} b(\bpi)  \label{def-X} \\
Z &=& \max_{\bpi \in \EE_2} b(\bpi)
- \max_{\bpi \in \EE_0} b(\bpi)  . \label{def-Z}
\end{eqnarray}
To see why such definitions are useful, note that 
the solution $\bpi$ to (\ref{inf-prob})
will have a path through the root if and only if
\[ \max_{\bpi \in \EE_2} b(\bpi)
> \max_{\bpi \in \EE_0} b(\bpi) , \] 
that is if and only if $Z>0$.

We now set up the recursion that determines the joint
distribution of $(X,Z)$.
We remark that $X$ is introduced only because
it arises in the recursion for $Z$ -- it would obviously
be preferable to find a recursion involving only a single
quantity like $Z$, but that seems impossible to find.
Figure 4 may be helpful in visualizing the argument below.

By the recursive structure of the PWIT, 
for each subtree $(\bT_i, \ i = 1,2,3,\ldots)$
defined by the children of the root, the random pairs
$(X_i,Z_i)$ defined as at (\ref{def-X},\ref{def-Z})
on $\bT_i$ are distributed as $(X,Z)$
and are independent as $i$ varies.
We will first show
\begin{equation}
X = \max_i 
(\lambda - \xi_i + X_i - Z_i^+)
\label{X-rec}
\end{equation}
where $Z^+ = \max(0,Z)$
and where $\xi_i$ are the edge-lengths at the root.

Consider the families
$\bpi_1$ and $\bpi_0$
attaining the maxima over $\EE_1$ and $\EE_0$ in the definition 
(\ref{def-X}) of $X$.
So $\bpi_1$ contains an edge from the root to child $i$, say.
On the subtrees $(\bT_j, j \neq i)$ the maximal 
families must be identical, so we only need compare
$\bpi_1$ and $\bpi_0$ on the root-edges and the subtree $\bT_i$.
There is a contribution 
$\lambda - \xi_i$ to $b(\cdot)$ from the edge $(\root, i)$.
In the subtree $\bT_i$, we have
\begin{eqnarray*}
X_i &=& \max_{\bpi \in \EE_1(i)} b(\bpi)
- \max_{\bpi \in \EE_0(i)} b(\bpi)  \\ 
Z_i^+ &=& \max_{\bpi \in \EE_2(i) \cup \EE_0(i)} b(\bpi)
- \max_{\bpi \in \EE_0(i)} b(\bpi)  . 
\end{eqnarray*}
The family $\bpi_1$ contains the first-term maximizing family $\bpi
 \in \EE_1(i)$
in this definition of $X_i$,
while the family $\bpi_0$ contains the first-term maximizing family $\bpi
 \in \EE_2(i) \cup \EE_0(i)$
in this definition of $Z_i^+$.
So the contribution to $b(\bpi_1)$ from $\bT_i$
equals
$X_i - Z_i^+$.
This establishes (\ref{X-rec}), since we can choose the maximizing
value of $i$ to be the edge at the root.

A similar argument leads to a recursion for $Z$.
A family $\bpi_2$ containing a path through the root
must contain two edges
$(\root,i)$ and $(\root, j)$, say.
The contribution to $b(\cdot)$, relative to using no edges at the root,
of using $(\root,i)$ equals
$\lambda - \xi_i + X_i - Z_i^+$.
Hence we get
\begin{equation}
Z = 
\max_i (\lambda - \xi_i + X_i - Z_i^+)
+
{\max_i}^{[2]} (\lambda - \xi_i + X_i - Z_i^+)
\label{Z-rec}
\end{equation}
where $\max_i^{[2]}$ denotes second maximum.
Equations (\ref{X-rec},\ref{Z-rec}) together
give a formula for $(X,Z)$ in terms of 
$(X_i,Z_i), i \geq 1$
and $(\xi_i, i \geq 1)$.
By the recursive structure of the PWIT, the
$(X_i,Z_i), i \geq 1$
are independent copies of $(X,Z)$.
Thus (\ref{X-rec},\ref{Z-rec}) constitute
a 
{\em recursive distributional equation}
(RDE) for the ``unknown" joint distribution $(X,Z)$.

\setlength{\unitlength}{0.25in}
\begin{picture}(26,14)
\put(-3,0){
\begin{picture}(12,12)
\thinlines
\put(0.5,0.5){\line(1,0){11}}
\put(0.5,0.5){\line(0,1){11}}
\put(11.5,11.5){\line(0,-1){11}}
\put(11.5,11.5){\line(-1,0){11}}
\put(6,6){\circle*{0.2}}
\put(6,5){\circle*{0.2}}
\put(6,5.8){\line(0,-1){0.6}}
\put(6.1,5.3){$\xi^{\emptyset}_1$}
\put(4,8){\circle*{0.2}}
\put(5.8,6.2){\line(-1,1){1.6}}
\put(4.7,6.6){$\xi^{\emptyset}_3$}
\put(8,7){\circle*{0.2}}
\put(11,5){\circle*{0.2}}
\put(5,4){\circle*{0.2}}
\put(5.8,4.8){\line(-1,-1){0.6}}
\put(7,2){\circle*{0.2}}
\put(10,1){\circle*{0.2}}
\put(4,2){\circle*{0.2}}
\put(5,1){\circle*{0.2}}
\put(5,3.8){\line(0,-1){2.6}}
\put(3,8){\circle*{0.2}}
\put(3.8,8){\line(-1,0){0.6}}
\put(1,10){\circle*{0.2}}
\put(2.8,8.2){\line(-1,1){1.6}}
\put(4,10){\circle*{0.2}}
\put(7,11){\circle*{0.2}}
\put(2,4){\circle*{0.2}}
\put(1,5){\circle*{0.2}}
\put(1.8,4.2){\line(-1,1){0.6}}
\put(2,2){\circle*{0.2}}
\put(2,3.8){\line(0,-1){1.6}}
\put(9,10){\circle*{0.2}}
\put(10,8){\circle*{0.2}}
\put(11,8){\circle*{0.2}}
\put(10.2,8){\line(1,0){0.6}}
\put(11,11){\circle*{0.2}}
\put(10.07,8.2){\line(1,3){0.86}}
\put(4.17,7.94){$\scriptstyle{3}$}
\put(8.16,6.84){$\scriptstyle{2}$}
\put(5.61,4.89){$\scriptstyle{1}$}
\put(10.61,4.9){$\scriptstyle{4}$}
\put(5.6,5.9){$\emptyset$}
\put(0.8,10){\line(-1,0){0.6}}
\put(0.8,5){\line(-1,0){0.6}}
\put(1.9,1.8){\line(-1,-2){0.9}}
\put(5.1,0.9){\line(1,-1){0.9}}
\put(11.1,4.9){\line(1,-1){1.1}}
\put(11.1,5.1){\line(1,1){1.1}}
\put(11.1,8){\line(1,0){0.9}}
\put(11.1,11){\line(1,0){0.9}}
\end{picture}
}
\put(11,0){
\begin{picture}(12,12)
\thinlines
\put(0.5,0.5){\line(1,0){11}}
\put(0.5,0.5){\line(0,1){11}}
\put(11.5,11.5){\line(0,-1){11}}
\put(11.5,11.5){\line(-1,0){11}}
\put(6,6){\circle*{0.2}}
\put(6,5){\circle*{0.2}}
\put(4,8){\circle*{0.2}}
\put(8,7){\circle*{0.2}}
\put(11,5){\circle*{0.2}}
\put(5,4){\circle*{0.2}}
\put(7,2){\circle*{0.2}}
\put(10,1){\circle*{0.2}}
\put(4,2){\circle*{0.2}}
\put(4.9,3.8){\line(-1,-2){0.8}}
\put(5,1){\circle*{0.2}}
\put(5,3.8){\line(0,-1){2.6}}
\put(3,8){\circle*{0.2}}
\put(3.8,8){\line(-1,0){0.6}}
\put(1,10){\circle*{0.2}}
\put(2.8,8.2){\line(-1,1){1.6}}
\put(4,10){\circle*{0.2}}
\put(4,8.2){\line(0,1){1.6}}
\put(7,11){\circle*{0.2}}
\put(2,4){\circle*{0.2}}
\put(1,5){\circle*{0.2}}
\put(1.8,4.2){\line(-1,1){0.6}}
\put(2,2){\circle*{0.2}}
\put(2,3.8){\line(0,-1){1.6}}
\put(9,10){\circle*{0.2}}
\put(10,8){\circle*{0.2}}
\put(11,8){\circle*{0.2}}
\put(10.2,8){\line(1,0){0.6}}
\put(11,11){\circle*{0.2}}
\put(10.07,8.2){\line(1,3){0.86}}
\put(5.6,5.9){$\emptyset$}
\put(0.8,10){\line(-1,0){0.6}}
\put(0.8,5){\line(-1,0){0.6}}
\put(1.9,1.8){\line(-1,-2){0.9}}
\put(4,1.8){\line(0,-1){1.8}}
\put(5.1,0.9){\line(1,-1){0.9}}
\put(11.1,4.9){\line(1,-1){1.1}}
\put(11.1,5.1){\line(1,1){1.1}}
\put(11.1,8){\line(1,0){0.9}}
\put(11.1,11){\line(1,0){0.9}}
\put(4.1,10.2){\line(1,2){0.9}}
\put(4.17,7.94){$\scriptstyle{3}$}
\put(8.16,6.84){$\scriptstyle{2}$}
\put(5.61,4.89){$\scriptstyle{1}$}
\put(10.61,4.9){$\scriptstyle{4}$}
\end{picture}
}
\end{picture}

{\bf Figure 4.}
{\small 
On the realization of the PWIT from Figure 3, the left side
illustrates the optimal $\bpi \in \EE_2$ which does pass through the root
(which happens to use the edges from the root to $1$ and to $3$),
and the right side
illustrates the optimal $\bpi \in \EE_0$ which does not pass through the root.
These path-families coincide on the subtrees of children except
$\{1,3\}$.
On the subtree $\bT_3$, the optimal family on the right side
has a path through the root $3$, whereas on the subtree
$\bT_1$ it does not.
}

\vspace{0.12in}

We next show how the desired quantities
$\delta(\lambda)$ and $\eps(\lambda)$
at (\ref{del-lam},\ref{eps-lam})
can be obtained from the distribution of $(X,Z)$.
The quantity $\delta(\lambda)$ represents the proportion of
vertices in the optimal solution to (\ref{inf-prob}).
By the stationarity property of the root of the PWIT,
$\delta(\lambda)$ is just the probability that the optimal
family contains a path through the root.
As observed above, this happens if and only if $Z>0$, so
\begin{equation}
\delta(\lambda) = P(Z>0) .
\label{eq-delta}
\end{equation}
When $Z>0$, the lengths of the two edges in the path 
at the root are $\xi_I$ and $\xi_J$, where in the notation
of (\ref{Z-rec})
\begin{eqnarray*}
I &=& \arg \max_i (\lambda - \xi_i + X_i - Z_i^+)
\\
J &=& \arg {\max_i}^{[2]} (\lambda - \xi_i + X_i - Z_i^+)
 . \end{eqnarray*}
Again by stationarity,
the mean edge-lengths over \underline{all} edges in the optimal family
equals the mean edge-length in the edges \underline{at the root}
in the optimal family,
conditioned on the root being used, and so
\begin{equation}
\eps(\lambda) = \frac{E \left[ (\sfrac{\xi_I+\xi_J}{2}) \ 1_{(Z>0)} 
\right]}{\delta(\lambda)} .
\label{eq-eps}
\end{equation}

As mentioned before,
equations (\ref{X-rec},\ref{Z-rec}) together
form a {\em recursive distributional equation}
(RDE)
for the joint distribution of $(X,Z)$.
Such RDEs are pervasive not only in problems within $\SMF$
but also in many other areas of applied probability: see
\cite{me107} for a survey.
They rarely allow explicit solutions, but there is a standard
{\em bootstrap Monte Carlo} method 
(\cite{me107} section 8.1)
which is very easy to implement and which gives, in principle,
arbitrarily-accurate approximate solutions of RDEs.
This method was used to solve the RDE for $(X,Z)$
and then estimate
$\delta(\lambda)$ and $\eps(\lambda)$ via 
(\ref{eq-delta},\ref{eq-eps}).
Numerical values were shown in Table 1 and Figure 1.

\section{The connected edge-set case}
\label{sec-edge}
The conceptual ideas behind the analysis of 
$\tilde{\eps}(\delta)$ at (\ref{def-epstil})
are very similar to the analysis of
$\eps(\delta)$ in the previous section, so we will only
detail the differences.

Consider a forest
$\ff = (\bt_1,\bt_2,\ldots)$
in the PWIT, each of whose tree-components $\bt_i$ is infinite.
Define\\
\ \ 
$\FF$ is the set of such forests $\ff$;\\
\ \ 
$\FF_0$ is the set of such forests such that the root 
\underline{is not}
in any component;\\
\ \ 
$\FF_1$ is the set of such forests such that the root 
\underline{is}
in some component;\\
\ \ 
$\FF_2$ is the set of such forests, 
where in addition to the infinite tree-components we allow the
tree-component containing the root to be either empty, or finite, 
or infinite.

\vspace{0.09in}
\noindent
In the limit procedure which takes $\SMF$
to the PWIT, the limits of 
``connected edge-sets of size order $n$"
is exactly the set 
$\FF$ of forests whose tree-components are all infinite.
For $\ff = (\bt_i) \in \FF_2 \supset \FF = \FF_0 \cup \FF_1$,
consider
\[
b(\ff) = \lambda \times 
\#\{e: e \mbox{ an edge of some } \bt_i\} 
- \sum_{e: e \mbox{ edge of some } \bt_i} \xi_e
 . \]
The optimization problem on the PWIT is
\begin{equation}
\mbox{ maximize } b(\ff) \mbox{ over } 
\ff \in \FF . 
\label{inf-prob2}
\end{equation}
To study this we define
\begin{eqnarray}
Y &=& \max_{\ff \in \FF} b(\ff)
- \max_{\ff \in \FF_0} b(\ff)  \label{def-Y2} \\
Z &=& \max_{\ff \in \FF_1} b(\ff)
- \max_{\ff \in \FF_0} b(\ff) \label{def-Z2}\\
X &=& \max_{\ff \in \FF_2} b(\ff)
- \max_{\ff \in \FF_0} b(\ff)  . \label{def-X2}
\end{eqnarray}
Because
$  \FF_2 \supset \FF = \FF_0 \cup \FF_1$
we have
\[ X \geq Y = Z^+ . \]
The recursion for $X$, analogous to (\ref{X-rec}), is
\begin{equation}
X = \sum_i 
(\lambda - \xi_i + X_i - Y_i)^+
 .
\label{X-rec2}
\end{equation}
The argument is the same as for (\ref{X-rec}): the
contribution to $b(\cdot)$ by using edge $(\root,i)$, as compared
to not using it, equals
$(\lambda - \xi_i + X_i - Y_i)$,
and we may use any number, or zero, such edges.
The recursion for $Z$ is
\begin{equation}
Z = \max_I \left(
\sum_{i \in I}
(\lambda - \xi_i + Z_i - Y_i)
+ \sum_{i \not\in I} 
(\lambda - \xi_i + X_i - Y_i)^+
\right)
  \label{Z-rec2}
\end{equation}
where $I$ denotes a {\em non-empty} subset of
$\{1,2,3,\ldots\}$.
Here the first sum represents the contribution from the set
$I$
of children $i$ such that, in the optimal $\ff \in \FF_1$,
in the subtree $\bT_i$ the root $i$ is in an infinite component.
The set $I$ must be non-empty in order for $\ff \in \FF_1$.
Now the fact $X_i \geq Z_i$ implies
\[ \lambda - \xi_i + Z_i - Y_i \leq
(\lambda - \xi_i + X_i - Y_i)^+
\]
which implies there is an optimal $I$
with only one element, and we can rearrange 
(\ref{Z-rec2}) to become
\[ Z = X + \max_i
\left(
(\lambda - \xi_i + Z_i - Y_i)
-
(\lambda - \xi_i + X_i - Y_i)^+
\right)
 . \]
Finally, since $Y_i = Z_i^+$ we obtain the following RDE
for the joint distribution of $(X,Z)$.
\begin{eqnarray}
X &=& \sum_i 
(\lambda - \xi_i + X_i - Z_i^+)^+
\label{X-rec3}\\
Z &=& X + \max_i
\left(
(\lambda - \xi_i + Z_i - Z_i^+)
-
(\lambda - \xi_i + X_i - Z_i^+)^+
\right) .
 \label{Z-rec3}
\end{eqnarray}
We next show how the desired quantities
$\tilde{\delta}(\lambda)$ and $\tilde{\eps}(\lambda)$
can be obtained from $(X,Z)$.
Consider the optimal $\ff$ in
(\ref{inf-prob2}).
This $\ff$ contains the root if and only if $\ff \in \FF_1$,
that is if and only if $Z>0$, so
\begin{equation}
\tilde{\delta}(\lambda) = P(Z>0) .
\label{eq-delta2}
\end{equation}
When $Z>0$, the set ${\cal I}$ of edges at the root used in $\ff$
is the set of $i$ for which the contribution
$(\lambda - \xi_i + X_i - Z_i^+)$
is strictly positive, plus (if distinct) the maximizing $i$ in 
(\ref{Z-rec3}).
This leads to
\begin{equation}
\tilde{\eps}(\lambda) = \frac{
E \left[ 
(\frac{1}{2} \sum_{i \in {\cal I}} \xi_i)
\ 1_{(Z>0)}
\right] }
{\tilde{\delta}(\lambda)}
\label{eq-eps2}
\end{equation}
for ${\cal I}$ as above.

\section{Trees}
\label{sec-tree}
Studying trees $\bt$ in order to study the limit function
$\eps^*(\delta)$ at 
(\ref{def-eps*}) is a little more subtle.
What are the feasible solutions on the PWIT corresponding to
the limits of trees in $\SMF$?
At first sight they are just the set $\FF$ of forests
$\ff = (\bt_i)$ in section \ref{sec-edge}.
But this is wrong; instead, by analogy with many other
examples of limits of infinite trees
\cite{me52,LPS03}
the relevant feasible solutions are forests
$\ff = (\bt_i)$ with the extra property that each of whose tree-components $\bt_i$
have {\em one end};
that is, from each vertex of $\bt_i$ there is exactly one infinite
path in $\bt_i$.

To mimic the analysis of the previous section with 
this family of forests, it turns out we need, in place of
$\FF_2$ before, the family
defined as

$\FF_2$ is the set of such forests, modified so that the 
tree-component containing the root may be either empty or finite, 
but not infinite.

But now the analog of $X$ at (\ref{def-X2}) cannot be represented
recursively, since (roughly speaking) there is no recursive criterion
for finiteness.
Instead we 
need to consider, separately for $m = 0,1,2,\ldots$,
a definition such as

$\FF_{(m)}$ is the set of such forests, modified so that the 
tree-component containing the root  
has exactly $m$ edges.

Defining $X_m$ in terms of a maximum over $\FF_{(m)}$ leads
to a RDE for the infinite family
$(X_0,X_1,X_2,\ldots,Z)$.
But we have not attempted to solve this numerically.

Fortunately, this detailed analysis is unnecessary for
investigating the scaling exponent because 
\[
\eps^*(\delta) = \tilde{\eps}(\delta)
\mbox{ when } 
\tilde{\eps}(\delta) < e^{-1} . \]
To outline the argument, consider the minimizing edge-set $\be$
for $\tilde{\eps}(\delta)$ in this range.
Suppose $\be$ contains a cycle of length order $n$.
By the fact (for {\em paths}) $\eps(0+) = e^{-1}$, this cycle has average
edge-length $> e^{-1}$ and hence has some edge of length
$> e^{-1}$.
Removing this edge would reduce $A(\be)$ without essentially affecting the constraint on $\len(\be)$, contradicting minimality.
So $\be$ can have no cycles of length order $n$.
As for short cycles, fix $a < e^{-1}$ and consider a typical point
$v$ of $\SMF$.
By the arguments of \cite{me75,janson-123}
(comparison with the Yule process),
\[ P(v \mbox{ in any cycle } \mathbf{c} \mbox{ with } 
A(\mathbf{c}) < a) \to 0
\mbox{ as } n \to \infty \]
and it follows that the contribution to $A(\be)$ from short cycles
$\to 0$ as $n \to \infty$.

\section{Final remarks}
\subsection{Sampling errors in Table 1}
\label{sec-sample}
We treat the case of trees; the case of paths could be treated similarly.
To obtain the numerical values in Table 1, 
we represented the distribution $(X,Z)$ via $10^6$ points
and iterated the RDE 1000 times, truncating the
Poisson process $(\xi_i, 1 \leq i < \infty)$
at $i = 20$.
This necessitated, for each value of $\lambda$, a total of
$2 \times 10^{10}$ calls to the random number generator.
We calculated $\eps$ and $\delta$ using the final 200 generations,
that is using $2 \times 10^8$ points.
There are various possible errors in this way of estimating scaling exponents,
of which the only one which can be quantified is ``sampling error".
Clearly
\[ \mbox{s.d. (estimate of } 
\delta \mbox{)}  \approx \delta^{1/2}/\sqrt{2 \times 10^8}
\approx 0.7 \times 10^{-4} \ \delta^{1/2} \]
which is negligible.
But the error for $\eps$ is not negligible,
since it is based on only a proportion $\delta$ of the samples,
giving
\[ \mbox{s.d. (estimate of } 
\eps \mbox{)} 
\approx \frac{\mbox{s.d.($\xi$)}}{\sqrt{2 \times 10^8 \ \delta}} \]
where s.d.($\xi$) $\approx 0.3$ is the s.d. of the $\xi$-values used
to estimate $\eps$
via (\ref{eq-eps2}).
This leads to
\[ \mbox{s.d. (estimate of } 
\eps/\delta^{1/2} \mbox{)}
\approx 2 \times 10^{-5} \ \delta^{-1} \]
which are the $\pm$ values shown in Table 1.

\subsection{Rigorous bounds on scaling exponents}
Because the limit $\eps(0+) = e^{-1}$ in the paths setting
is essentially just a first moment calculation,
a referee suggests that similar first moment methods should
establish rigorously some bound on $\eps(\delta)$ and hence
some bound of the scaling exponent in the paths case.
We concur, but have not attempted a detailed calculation.

\subsection{Scope of scaling exponents}
\label{sec-scope}
It seems difficult to specify precise the range of settings
in which a definition of {\em percolation-like scaling exponent}
makes sense and is interesting.
Within the stochastic mean field model there is a well studied
{\em minimum matching} problem 
(see \cite{LW03,NPS03} for recent proofs of the Parisi conjecture)
in which context one could define
\[
\eps^{\mbox{\tiny match}}(\delta)
= \lim_n  E
\frac{\mbox{\small length min matching of some $\delta n$ vertices}}
{\frac{1}{2} \delta n}
. \]
But here it is clear that
\[
\eps^{\mbox{\tiny match}}(\delta)
\sim \delta \mbox{ as } \delta \downarrow 0 \]
so that the critical value equals $0$ and the scaling
exponent equals $1$.
However, since the critical value equals zero we are inclined to
regard this case as ``not percolation-like".
A referee suggests the example 
(again, within the stochastic mean field model) 
of the path through $\delta n$ points chosen greedily by
choosing the shortest available edge at each successive vertex,
but this also seems ``not percolation-like".

\paragraph{Acknowledgement.}
I thank two anonymous referees for helpful comments.


\def\cprime{$'$} \def\polhk#1{\setbox0=\hbox{#1}{\ooalign{\hidewidth
  \lower1.5ex\hbox{`}\hidewidth\crcr\unhbox0}}} \def\cprime{$'$}
  \def\cprime{$'$} \def\cprime{$'$}
  \def\polhk#1{\setbox0=\hbox{#1}{\ooalign{\hidewidth
  \lower1.5ex\hbox{`}\hidewidth\crcr\unhbox0}}} \def\cprime{$'$}
  \def\cprime{$'$} \def\polhk#1{\setbox0=\hbox{#1}{\ooalign{\hidewidth
  \lower1.5ex\hbox{`}\hidewidth\crcr\unhbox0}}} \def\cprime{$'$}
  \def\cprime{$'$} \def\cydot{\leavevmode\raise.4ex\hbox{.}} \def\cprime{$'$}
  \def\cprime{$'$} \def\cprime{$'$} \def\cprime{$'$}

\end{document}